\documentclass{article}
\usepackage{lineno}
\usepackage{xcolor}
\usepackage{amsmath}

\usepackage{PRIMEarxiv}

\usepackage[utf8]{inputenc} 
\usepackage[T1]{fontenc}    
\usepackage{hyperref}       
\usepackage{url}            
\usepackage{booktabs}       
\usepackage{amsfonts}       
\usepackage{nicefrac}       
\usepackage{microtype}      
\usepackage{lipsum}
\usepackage{fancyhdr}       
\usepackage{graphicx}       
\graphicspath{{media/}}     

\usepackage{authblk}
\usepackage{parskip}
\usepackage[absolute]{textpos}

\title{Development of High Granular Neutron Time-of-Flight Detector for the BM@N experiment}

\author[*,1]{F. Guber}
\author[1]{D. Finogeev}
\author[1]{M. Golubeva}
\author[1]{A. Ivashkin}
\author[1]{A. Izvestnyy}
\author[1]{N. Karpushkin}
\author[1,3]{A. Makhnev}
\author[1,2]{D. Lyapin}
\author[1,2]{\\M. Mamaev}
\author[1]{S. Morozov}
\author[1,2]{P. Parfenov}
\author[1]{A. Shabanov}
\author[1]{D. Serebryakov}
\author[1,2]{A. Zubankov}

\affil[1]{Institute for Nuclear Research of the Russian Academy of Sciences, Moscow, Russia}
\affil[2]{National Research Nuclear University MEPhI, Moscow, Russia}
\affil[3]{Moscow Institute of Physics and Technology, Dolgoprudny, Russia}

\begin{document}
\maketitle

\begin{abstract}
The HGND (High Granular Neutron Detector) is developed for the BM@N (Baryonic Matter at Nuclotron) experiment on the extracted beam of the Nuclotron at JINR, Dubna. The HGND will be used to measure the azimuthal flow of neutrons produced with energies ranging from 300 to 4000 MeV in heavy-ion collisions at beam energies of 2--4 AGeV. The azimuthal flow of charged particles will be measured using the BM@N magnet spectrometer. The data on the azimuthal flow of neutrons will shed light on the study of the high-density Equation of State (EoS) of isospin-symmetric nuclear matter, which is crucial for studying astrophysical phenomena such as neutron stars and their mergers. The HGND has a highly granular structure with approximately 2000 plastic scintillation detectors (cells), each measuring 4$\times$4$\times$2.5 cm$^3$. These detectors are arranged in 16 layers, with 121 detectors in each layer, and are subdivided by copper absorber plates with a thickness of 3 cm. The light from each cell is detected with SiPM (Silicon Photomultiplier) with an active area of 6$\times$6 mm$^2$. Developed multi-channel TDC board based on the Kintex FPGA chip with a bin width of 100 ps will be used to perform precise timestamp and amplitude measurement using Time-over-Threshold (ToT) method. Good spatial resolution due to the high granularity together with a cell's time resolution of 100-150 ps ensures neutron reconstruction with good energy resolution. The design of the detector as well as the results from test measurements and simulations have been presented.
\end{abstract}

\begin{textblock*}{\textwidth}(2.5cm,26cm)
  \begin{flushleft}
    \noindent\rule{7cm}{0.4pt} \\
    * Corresponding author. Email: guber@inr.ru
  \end{flushleft}
\end{textblock*}

\section{Introduction}
\label{sec:intro}

The study of the equation of state (EoS) of dense nuclear matter \cite{Sorensen:2023zkk} is one of the central topics in contemporary nuclear physics. The EoS establishes the relationship between pressure, density, energy, temperature, and the symmetry energy. The symmetry energy term characterizes the isospin asymmetry of nuclear matter: \[\delta = \left(\rho_n - \rho_p\right)/\rho,\] where $\rho_n$, $\rho_p$, and $\rho$ are the densities of neutron, proton, and nuclear matter, respectively. Constraining this term is very important for astrophysics because the mass and radii relation of neutron stars and the dynamics of neutron star mergers strongly depend on the contribution of the symmetry energy term in the EoS of high-density neutron matter \cite{Senger:2021dot}. 

In the last two decades, the most stringent constraints on the symmetric nuclear matter EoS have come from the available measurements of anisotropic flow of charged particles in isospin-symmetric heavy-ion collision experiments. Anisotropic flow is expressed in the coefficients of the Fourier decomposition of the azimuthal distribution of the particles produced in the collision with respect to the reaction plane (the plane spanned by the vectors of the impact parameter and beam direction): \[\rho(\phi) = \frac{1}{2\pi}\left(1+2\sum_{n=1}^{\infty}v_n (cos(n(\phi-\Psi_{RP})))\right),\] where $\phi$ is the azimuthal angle of the particle produced in the collision, $\Psi_{RP}$ is the reaction plane angle, and \[v_n = \langle \cos n(\phi-\Psi_{RP}) \rangle.\] The first and the second coefficients of the decomposition, $v_1$ and $v_2$, are called directed flow and elliptic flow, respectively.

Most of the data have been obtained in Au+Au collisions, performed at beam energies up to 1.5A GeV at GSI (Darmstadt, Germany) with beams from SIS18, and at BNL (USA) using beams from AGS. At SIS18, ions of gold were accelerated up to kinetic energies of 1.5 AGeV, where average nuclear densities reach up to 2 times the nuclear density $\rho_0$. The FOPI collaboration at GSI measured the elliptic flow $v_2$ of protons, deuterons, tritons, and 3He in Au + Au collisions at beam kinetic energies from 0.4 AGeV to 1.5 AGeV \cite{LeFevre:2015paj}. It was shown that a moderately soft EoS describes the data of elliptic flow $v_2$ of protons rather well, although some discrepancies were observed between different sets of experimental data. The ratio of the directed and elliptic neutron flow to the corresponding flow of protons, $v_{1,2}^n / v_{1,2}^{ch}$, is also a sensitive observable of the symmetry energy contribution to the EoS of high-density nuclear matter. Currently, only a few experimental results on the study of the ratio of the directed and elliptic neutron flow to corresponding flow of charged particles with Z=1 $v_{1,2}^n / v_{1,2}^{ch}$ obtained in Au + Au collisions at energies of 400, 600, and 800 MeV per nucleon in FOPI / LAND experiments [\cite{FOPI:1993wdf}, \cite{FOPI:1994xpn}] exist. The most reliable data were obtained later in the ASY EOS experiment with significantly better accuracy but only at an energy of 400 MeV per nucleon \cite{Russotto:2011hq}. It was shown that the elliptic flow ratio as a function of transverse momentum, $v_2^n (p_T) / v_2^{ch} (p_T)$, is consistent with a moderately soft EoS, although this estimation of the symmetry term suffers from noticeable systematic errors. At higher energies of 600 and 800 MeV per nucleon, the probability of hadron showers in the LAND neutron detector \cite{LAND:1991ffr} increases, and the reliability of the obtained data decreases. Currently, a new experiment is proposed at SIS18 to measure the symmetry energy in Au+Au collisions at energies of 400, 600, and 1000 MeV \cite{Russotto:2021mpu}. A part of the new neutron detector NeuLAND \cite{R:2021lxa}, developed for the R3B experiment at FAIR, will be used. The soft term of the dense nuclear matter EoS is also confirmed in kaon production experiments in nucleus-nucleus collisions at SIS18 energies \cite{Senger:2022ihi}.

The directed and elliptic flow of protons at higher nuclear densities have been measured in Au+Au collisions at beam kinetic energies between 2 AGeV and 11 AGeV at the AGS in Brookhaven \cite{E895:1999ldn, E895:2000maf}. Here, it was shown that the elliptic flow of protons at 2 AGeV is more compatible with a hard EOS \cite{E895:2001axb}. On the other hand, astrophysical experimental measurements of the masses of the most massive neutron stars with high nuclear densities exclude a soft EOS \cite{Sorensen:2023zkk}. New precise measurements are planned for both directed and elliptic flow in heavy-ion collisions to further constrain the EOS of high-density nuclear matter at future experiments at NICA and FAIR facilities, where baryon densities of more than 5 times the nuclear density $\rho_0$ can be reached, which is close to the density in the core of neutron stars \cite{articleSenger}.

Good opportunities are now available to study the EoS at baryon densities of nuclear matter with (2-4) times the nuclear density $\rho_0$ and to obtain new experimental results at the BM@N (The Baryonic Matter at Nuclotron) --- the first running fixed-target experiment at the NICA facility \cite{Senger:2022bzm}. First Xe+CsI data taking at 3.8 AGeV was performed in 2023. In the future, heavier colliding systems at ion energies up to 4 AGeV will be studied. To measure the yields and flow of neutrons at the BM@N, a new high-granular time-of-flight neutron detector (HGND) is now being developed and constructed.

This article contains a short description of the BN@N setup and the position of HGND at the BM@N, design details with proposed electronics, as well as experimental results of time-of-flight resolution measurements of scintillation detectors (cells). Simulation results of the acceptance, neutron detection efficiency, neutron energy resolution, and neutron count rate estimations for the HGND are presented for the reaction Bi+Bi at 3 AGeV kinetic energy.

\section{The BM@N experiment setup}
\begin{figure}[htbp]
  \centering
  \includegraphics[width=.8\textwidth]{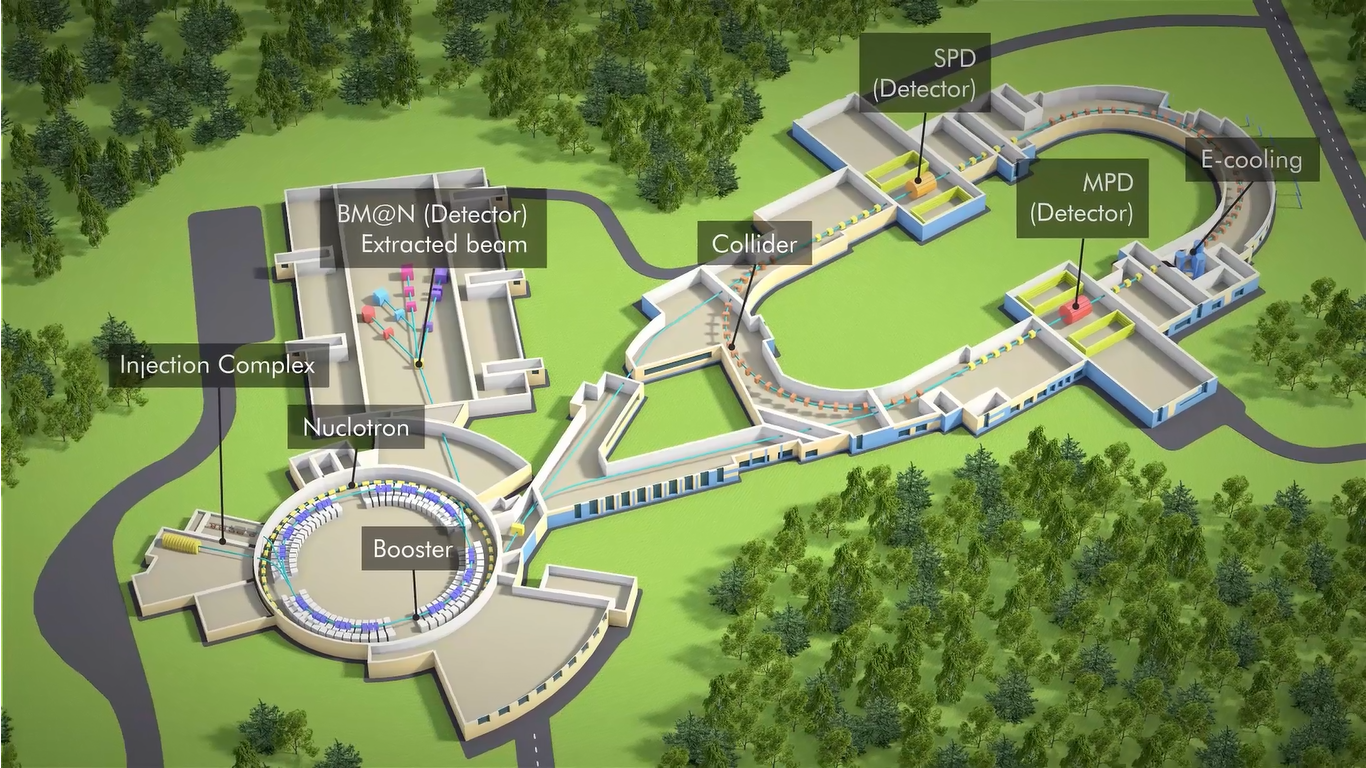}
  \caption{NICA facility complex at JINR, Dubna. The BM@N setup is placed on the extracted beam of the Nuclotron.}
  \label{fig:nica}
\end{figure}

\begin{figure}[htbp]
  \centering
  \includegraphics[width=.8\textwidth]{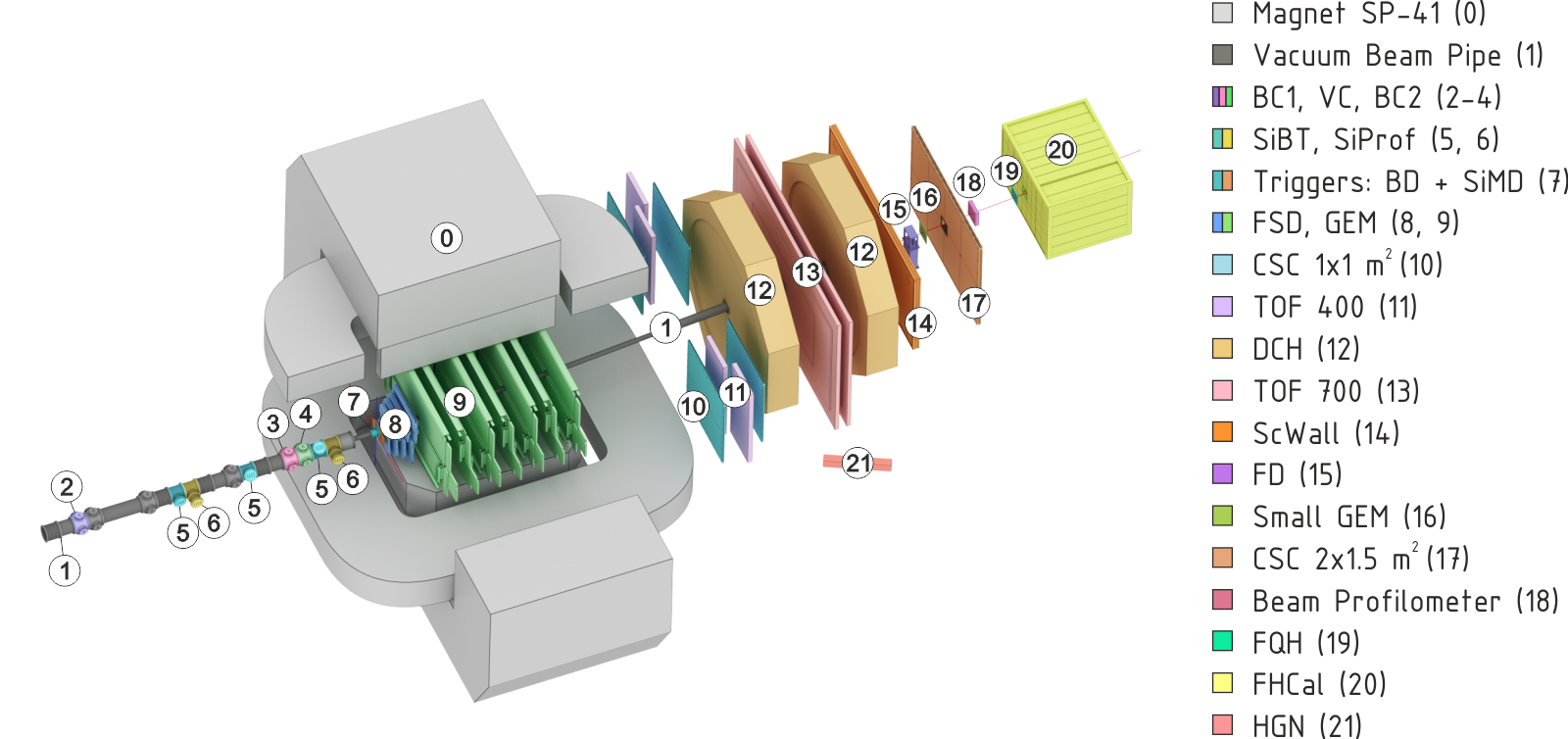}
  \caption{Schematic view of the BM@N setup in Xe+CsI run period.}
  \label{fig:setup}
\end{figure}

The BM@N experiment is part of the NICA complex (see Fig. \ref{fig:nica}), located on the extracted beam from the Nuclotron in the target hall \cite{Kapishin:2020cwk}. The Nuclotron provides beams of various particles, from protons to gold ions, with kinetic energy ranging from 1 to 6 GeV/nucleon for light ions with $Z/A\sim$ 0.5 and up to 4.5 GeV/nucleon for heavy ions with $Z/A\sim$ 0.4. Recently, in early 2023, the physics run at BM@N was conducted to study the Xe+CsI reaction with two kinetic energies of xenon ions: 3.0 and 3.8 AGeV. The BM@N setup (see Fig. \ref{fig:setup}) comprises a normal conducting dipole magnet with a gap of 96 cm and a pole length of about 1 m, along with several detector systems to identify and measure the momentum of produced charged particles, as well as the geometry of nucleus-nucleus collisions. The magnetic field can be adjusted to up to 1.2 T to optimize the BM@N detector acceptance and momentum resolution for different colliding systems and beam energies.

Particle identification is performed by time-of-flight measurement using TOF-400 and TOF-700 placed at distances of 4 m and 7 m from the target, respectively. They are based on multi-gap resistive plate chambers (RPC) with strip read-out and provide a time resolution of 80 ps for TOF-400 and 115 ps for TOF-700. These detector systems allow the identification of hadrons ($\pi, K, p$) as well as light nuclei with momenta up to a few GeV/c produced in multi-particle events. The tracking system consists of three planes of two-coordinate Silicon detectors (STS) and six tracking stations based on triple Gas-Electron-Multiplier (GEM) chambers installed in the magnet, as well as two Cathode-Drift-Chambers (CDC) located behind the magnet. The STS + GEM system measures the momentum (p) of charged particles with a relative uncertainty that varies from 2.5\% at a momentum of 0.5 GeV/c to 4.5\% at 3.5 GeV/c. This system also provides measurements of the multiplicity of the produced charged particles, which can be used as an additional estimator of the collision centrality. To measure centrality and the orientation of the reaction plane in nucleus-nucleus collisions, the Forward Hadron Calorimeter (FHCal) is used, together with the Forward Quartz Hodoscope (FQH) and the multichannel Scintillation Wall (ScWall), which is also used to measure charged spectator fragments. There are also some beam and trigger detectors listed in Fig. \ref{fig:setup}.

\section{Conception of the New High Granular Neutron Detector (HGND) at the BM@N}
To measure neutron yields and the ratio of azimuthal flow of neutrons and protons, it is necessary to develop a new detector for neutron identification and energy reconstruction. Up to now, only two neutron detectors exist to measure neutrons with energy up to 1 GeV: LAND \cite{LAND:1991ffr} and NeuLAND \cite{R:2021lxa}, both developed and constructed at GSI. These neutron detectors use the time-of-flight method to identify neutrons and measure their energy. The active part of these detectors consists of long scintillator plates with PMT light readout from both ends. LAND provides a time resolution of about 250 psec and an 80\% detection efficiency for a single neutron with energy above 400 MeV. In contrast to LAND, which has a structure with alternating layers of plastic scintillator and iron with thicknesses of 5 mm, NeuLAND consists of only plastic scintillator layers (without absorber) and has a total longitudinal length of 3 m instead of 1 m as for LAND. NeuLAND has a time resolution of about 150 psec and a 95\% detection efficiency for a single neutron in the energy range of 400 -- 1000 MeV. In the Au+Au reaction at 400 AMeV, the ratio of the flow of neutrons/charged particles with Z=1 was measured \cite{Russotto:2021mpu} using only LAND.

Taking into account the significantly higher energy range of neutrons (up to 4 GeV) produced in heavy-ion collisions at the BM@N and the strong hadron shower development at these energies in the detector volume, a new type of neutron detector (HGND) is proposed for development. Instead of long scintillator plates, it is proposed to use an array of small cells with individual light readout by SiPM. Each cell should provide a time resolution of 100-150 ps. Copper absorbers between active layers are needed to provide the required interaction length of the HGND in the limited available space at the BM@N. The details of the HGND design will be discussed in the next section.

\begin{figure}[htbp]
  \centering
  \includegraphics[width=.8\textwidth]{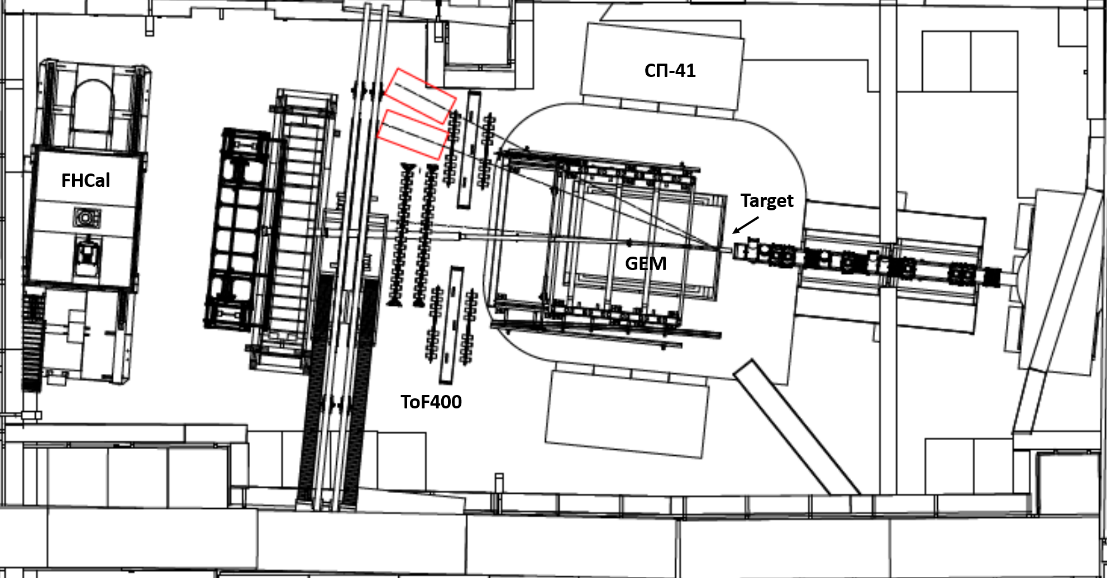}
  \caption{Positions of the HGND (red boxes) at neutron production angle 17.0 and 22.3 degrees.}
  \label{fig:positions}
\end{figure}

It is proposed to place the new neutron detector just behind TOF400 at a distance of 500 cm from the target to the entrance surface (red boxes in Fig. \ref{fig:positions}). At this distance, neutrons can be measured in the angle range of 17.0 - 22.3 degrees relative to the ion beam axis with the minimum material budget. Due to the limited available space, the length of the neutron detector cannot exceed 1 m.

\section{Design of the Neutron Detector}
The HGND consists of 16 alternating active layers with 121 cells grouped in an 11$\times$11 array, with copper absorber plates placed in between. The first scintillation layer will be used as a charged particle VETO detector. The absorber plates, each measuring 44$\times$44$\times$3 cm$^3$, and the scintillator layers are mounted on a common support frame. The total length of the HGND is about 1 m, corresponding to approximately 3 nuclear interaction lengths.

The cells are made up of 4$\times$4$\times$2.5 cm$^3$ plastic scintillators based on polystyrene with additions of 1.5\% paraterphenyl and 0.01\% POPOP. This plastic scintillator, with a light decay time of 3.9$\pm$0.7 ns, is produced at JINR. One of the large faces (40$\times$40 mm$^2$) of the scintillators will be covered with a black light-absorbing layer, leaving a 6$\times$6 mm$^2$ window in the center for the installation of a silicon photomultiplier. The other faces will be coated with a white dye based on $TiO_2$ to provide efficient diffuse reflection. The light read-out is performed with a silicon photomultiplier (SiPM) EQR15 11-6060D-S \cite{NDL-EQR15}. This SiPM has an active area of 6$\times$6 mm$^2$ with a 15$\times$15 $\mu m^2$ pixel size and a total number of pixels of 160,000. The SiPM has a quantum efficiency of 45\% and a gain of $4\times10^5$. As an option, the use of fast scintillators EJ-230 \cite{EJ-230} was also considered. The measured time resolution of the cell based on JINR scintillator is 117$\pm$2 ps \cite{Guber:2023ktn}. Due to the significantly faster light decay time of EJ-230 (2.8$\pm$0.5 ns), the measured time resolution of the detector based on EJ-230 is 74$\pm$1 ps \cite{Guber:2023ktn}. These time resolutions have been obtained using slewing correction \cite{Karpushkin:2023oqd}. Due to the high cost of EJ-230, the use of JINR scintillators is now considered as the main option for the HGND.

\begin{figure}[htbp]
  \centering
  \includegraphics[width=\textwidth]{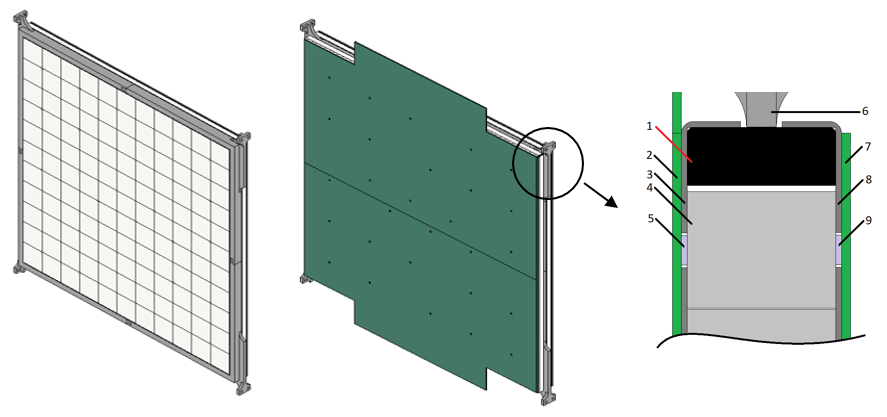}
  \caption{Schematic view of the assembly of the HGND scintillation layer. 
  Left: arrangement of 121 cells in the layer case. Center: view of layer with 2 PCBs attached to the scintillator array. Right: view of a transverse cross section of fully assembled scintillation layer: 1 -- the frame of the case; 2 -- one of the two PCBs with 55 or 66 SiPMs (5); 3 and 8 - aluminum plates for both sides of the frame case with cutouts for SiPMs and LEDs; 4 -- scintillator; 5 -- SiPM;  6 -- layer support bracket; 7 -- LED PCB; 9 -- LED;}
  \label{fig:HGN}
\end{figure}

Each active layer of the HGND consists of a 3D-printed PETG light-tight casing with 121 scintillators in an 11$\times$11 arrangement (Fig. \ref{fig:HGN}: 4). The casing is capped with two 1mm spacers made out of bent and stamped aluminum sheets (Fig. \ref{fig:HGN}: 3, 8). These sheets have circular cutouts, which are placed in front of each scintillator's center. Two PCBs with 55 and 66 SiPMs, respectively, are mounted on one side of the casing (Fig. \ref{fig:HGN}: 2), providing light readout and hosting the preamplifier and LVDS comparators, which generate ToT signals for the readout schematics. On the other side of the casing, a PCB with 121 LEDs (Fig. \ref{fig:HGN}: 7) is mounted, providing full-circuit calibration capabilities.

Due to the high total number of readout channels in the HGND (1936), a multi-channel readout system is developed based on the Kintex FPGA chip. The implemented TDCs perform precise timestamp and amplitude measurements using the Time-over-Threshold (ToT) method \cite{Finogeev_arxiv}. The bin width of every TDC channel is about 100 ps to meet the cell time resolution requirements. The dynamic range of signals from cells is relatively small, ranging from 1 to 10 MIP (Minimum Ionizing Particle).

\begin{figure}[htbp]
  \centering
  \includegraphics[width=.5\textwidth]{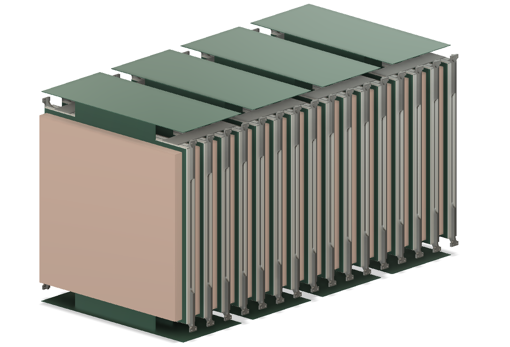}
  \caption{Schematic view of the assembled HGND with 4 readout boards on the top and below the detector.}
  \label{fig:HGN_elem}
\end{figure}

A single readout board with 3 FPGAs will host 250 channels, and eight boards in total will be necessary to readout about 2000 HGND channels. The readout boards carry all the heat-generating components, such as FPGAs, power supply ICs, and communication transceivers. Each of the readout boards interfaces with 4 "half-layers" via PCB edge connectors. A schematic view of the HGND arrangement is shown in Fig. \ref{fig:HGN_elem}. The FPGA-based read-out has time synchronization with the White Rabbit network and will be integrated into the BM@N common data acquisition system.

\section{Monte Carlo Simulation of the HGND}
\begin{figure}[htbp]
  \centering
  \includegraphics[width=0.9\textwidth]{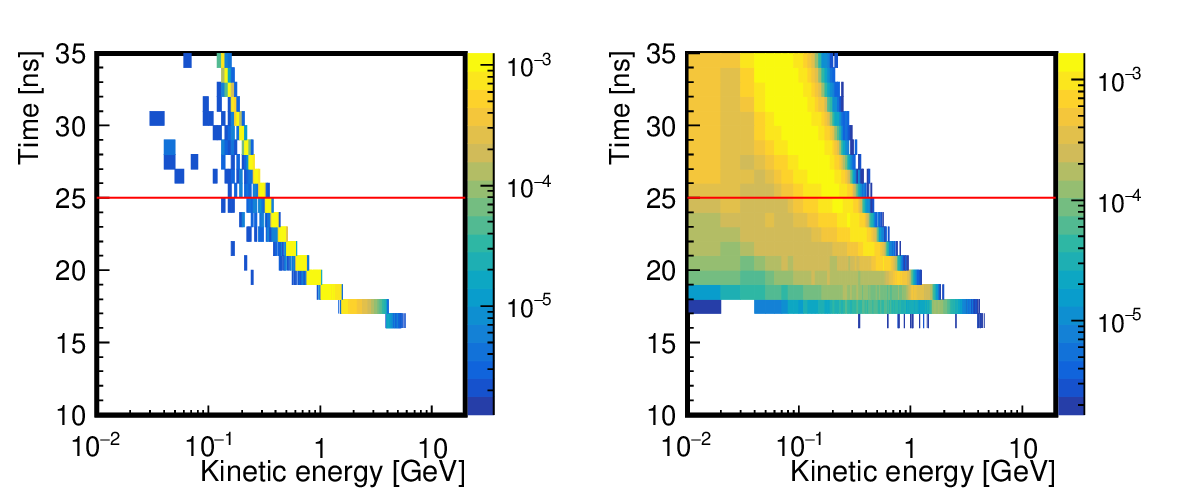}
  \caption{Left: the dependence of primary neutrons' time-of-flight on their kinetic energies at the entrance of the HGND. Right: Corresponding plot for background neutrons and all charged particles on the surfaces of the HGND.}
  \label{fig:tof_kin}
\end{figure}

Monte Carlo simulations of the HGND have been conducted to study acceptance, rapidity, transverse momentum distributions of neutrons, background conditions, etc. About $10^6$ minimum bias events for Bi + Bi reaction at 3.0 AGeV were produced with the DCM-QGSM-SMM event generator \cite{Amelin:1990js, Baznat:2019iom}. The propagation of particles through the BM@N and HGND was simulated using GEANT4 \cite{Brun:1994aa}, integrated into the BmnRoot software framework \cite{bmnroot}. The HGND was placed at an angle of 17 degrees relative to the ion beam axis and a distance of 500 cm from the target. The dependence of the primary neutron's time of flight on the entrance of the HGND versus its kinetic energy is shown in Fig. \ref{fig:tof_kin}, left. The same plot for background neutrons and all charged particles that hit the HGND volume from any side is shown in Fig. \ref{fig:tof_kin}, right. To reduce a significant part of the background, it is proposed to analyze only particles that arrive at the entrance of the HGND in the first 25 ns after the interaction at the BM@N target (horizontal lines on Fig. \ref{fig:tof_kin}). With this time cut, only neutrons with kinetic energy of more than 300 MeV will be selected. The number of background neutrons in this energy range will be suppressed by a factor of 6.

\begin{figure}[htbp]
  \centering
  \includegraphics[width=\textwidth]{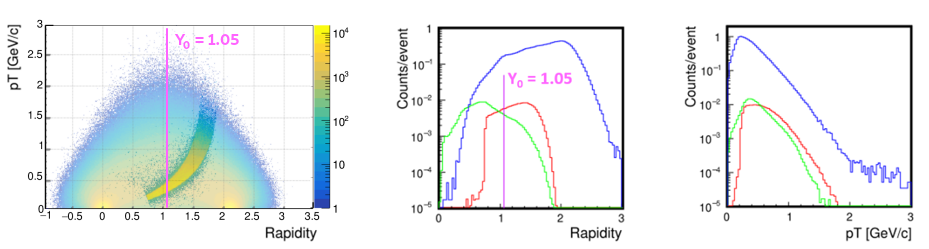}
  \caption{Left: the plot showing transverse momentum vs rapidity for primary neutrons at Bi+Bi@3AGeV. The plot illustrates the acceptance of the HGND at 17 degrees for primary neutrons with a time cut of 25 ns at the entrance of the HGND. The rapidity and transverse momentum distributions for primary (red color) and background (green color) neutrons are shown in the middle and right figures, respectively. The vertical line on the rapidity distribution corresponds to midrapidity ($y_0 = 1.05$) for the given reaction. The blue curves represent the corresponding distributions for the protons measured by the BM@N spectrometer.}
  \label{fig:acceptance}
\end{figure}

The plot illustrating transverse momentum vs. rapidity for primary neutrons in Bi+Bi@3AGeV is presented in Fig. \ref{fig:acceptance}, left. Here, the plot depicts the acceptance of the HGND for primary neutrons with a time cut of 25 ns at the entrance of the HGND. The rapidity and transverse momentum distributions for primary neutrons (red curve) and backgrounds (green curve) are displayed in Fig. \ref{fig:acceptance} (middle and right, respectively). It is evident that at this angle, the HGND accepts primary neutrons in the rapidity range of 0.8 to 1.8, including midrapidity ($y_0 = 1.05$), as indicated by the pink vertical line. For comparison, the rapidity and transverse momentum distributions of protons measured by the BM@N spectrometer are depicted by the blue curves.

\begin{figure}[htbp]
  \centering
  \includegraphics[width=\textwidth]{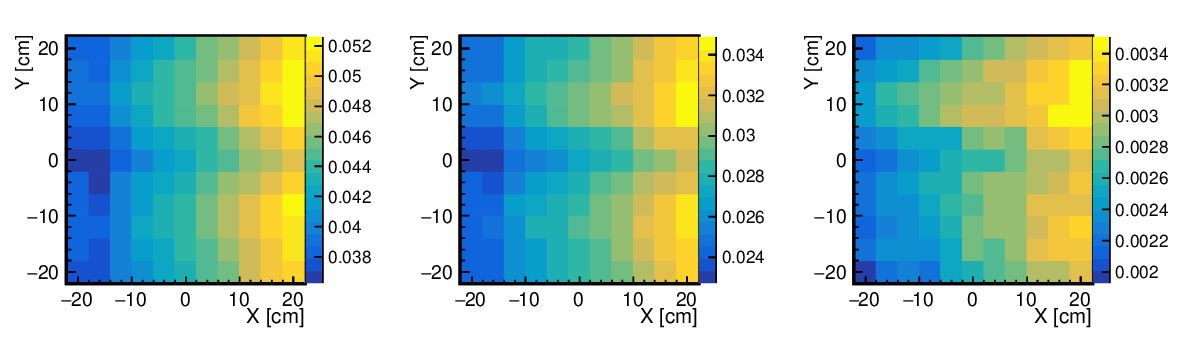}
  \caption{Left: Hit distributions of all particles in the scintillators of the VETO detector normalized to one interaction in the target. Middle: The same plot with a 25 ns time cut. Right: The probability of having 2 or more hits in the VETO scintillators.}
  \label{fig:hits}
\end{figure}

The left plot in Fig. \ref{fig:hits} displays hit distributions of all particles in cells of the VETO layer without a time cut. The middle plot in Fig. \ref{fig:hits} shows a similar distribution with a time cut of 25 ns. A higher hit rate is observed on the side of the VETO closer to the beam axis. The hit rates are normalized to one interaction in the target. At the maximum expected interaction rate of about 7 kHz, the hit rate in a single VETO cell is expected to be around 300 hits per second. Due to the low hit rates of the cells, the probability of having 2 or more hits in one scintillator is less than 0.3\% (Fig. \ref{fig:hits}, right).

\begin{figure}[htbp]
  \centering
  \includegraphics[width=\textwidth]{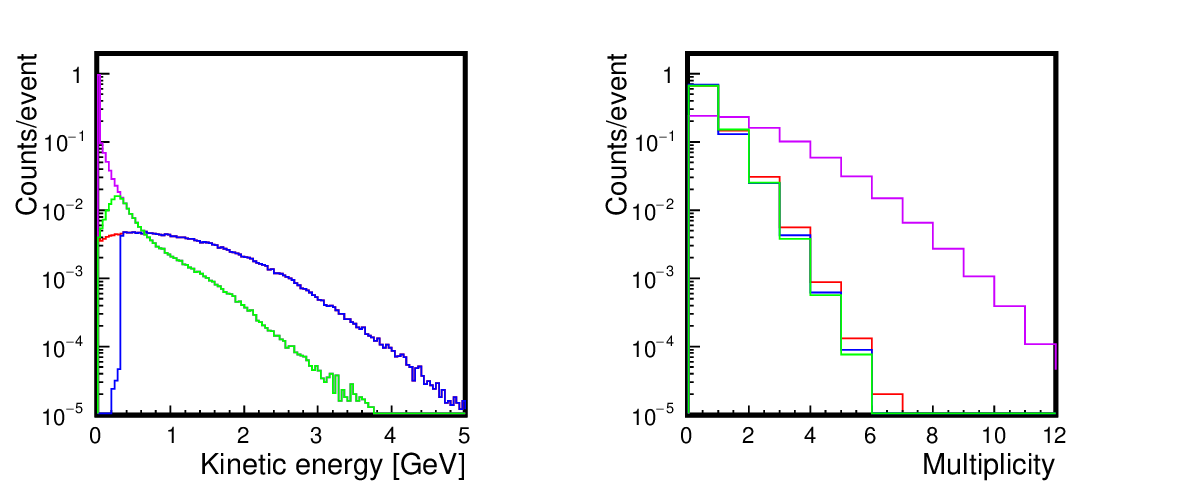}
  \caption{The neutron energy spectra (left) and multiplicity distributions (right) for primary (blue) and background (green) neutrons at the entrance surface of the HGND with a 25 ns time cut. The red and pink curves correspond to primary and background neutrons without a time cut.}
  \label{fig:energy_spectrum}
\end{figure}

The energy spectra and multiplicity distributions of primary (blue) and background (green) neutrons with a time-of-flight < 25 ns at the entrance surface of the HGND at 17.0$^{\circ}$ are shown in Fig. \ref{fig:energy_spectrum}. They are compared with the corresponding neutron energy spectra without a time cut (red and pink).

It can be seen that for primary neutrons with energies > 300 MeV, neutron multiplicity = 1 dominates. The use of a 25 ns time cut significantly reduces the multiplicity of background neutrons in the considered energy range of primary neutrons.

\begin{figure}[htbp]
  \centering
  \includegraphics[width=\textwidth]{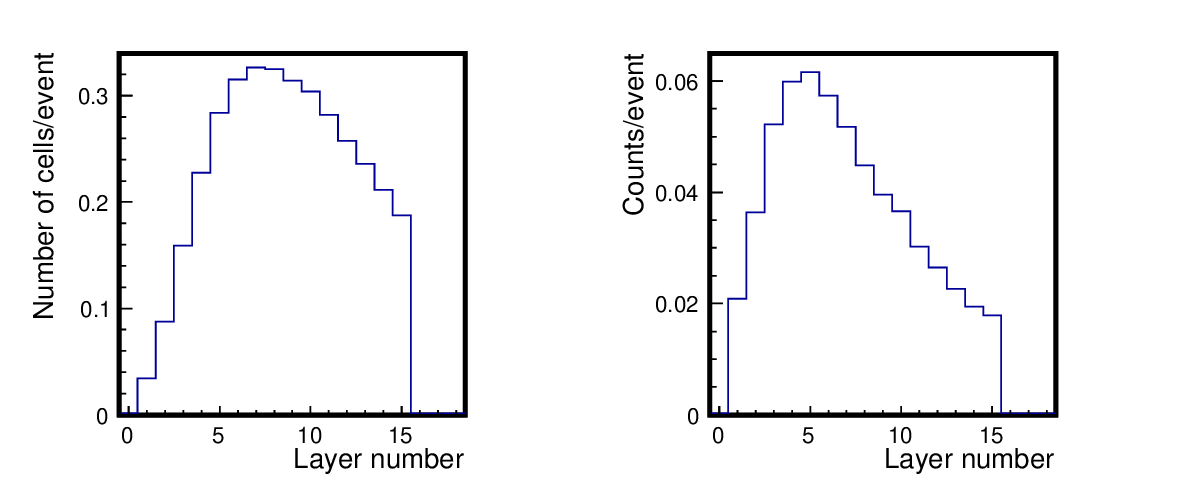}
  \caption{Left: The distribution of layers with the number of fired cells per event. Right: The distribution of layers with the first fired cell in the event.}
  \label{fig:layers_fired}
\end{figure}

At present, neutron identification and energy reconstruction in the HGN with realistic backgrounds and neutron multiplicities are under development. Nevertheless, the upper limit of the HGN detection efficiency can be estimated from simulated primary neutrons with multiplicity = 1. These neutrons propagated through the HGND and triggered some cells in the active layers of the neutron detector. The distribution of layers with the number of fired cells per event is shown in Fig. \ref{fig:layers_fired}, left. The distribution of layers with the first fired cell in the event is shown in Fig. \ref{fig:layers_fired}, right.

The simulated response times of cells were smeared according to a 150 ps time resolution. Only cells with energy deposition greater than 3 MeV (1/2 MIP) are taken into account. In order to reconstruct neutron energy, the time of the first fired cell was used.

\begin{figure}[htbp]
  \centering
  \includegraphics[width=\textwidth]{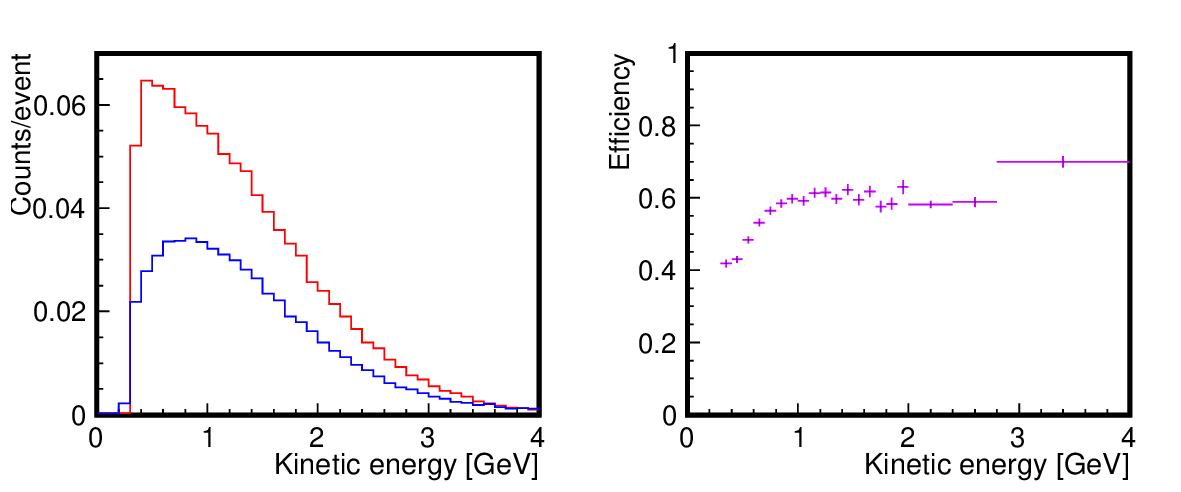}
  \caption{Left: The energy spectrum of primary neutrons with multiplicity = 1 (red) on the HGND surface and the reconstructed neutron spectrum (blue). Right: The efficiency of the energy reconstruction as a function of neutron kinetic energy.}
  \label{fig:primary_neutrons}
\end{figure}

Knowing the time and distance for a given cell from the target, the energy spectra of primary neutrons have been reconstructed, as shown in Fig. \ref{fig:primary_neutrons}, left. The energy spectrum for primary neutrons with multiplicity = 1 obtained from the event generator is shown in red. The corresponding reconstructed energy spectrum calculated with the time of the first fired cell is shown in blue. The efficiency of neutron reconstruction calculated from these spectra is shown as a function of neutron kinetic energy in Fig. \ref{fig:primary_neutrons}, right. It is necessary to emphasize that this efficiency should be considered as the upper limit. In reality, the efficiency will be lower because it will be necessary to take into account inefficiencies due to background rejection and the identification of primary neutrons with multiplicities greater than one. Methods based on determining clusters of the fired cells while considering energy depositions are under development. Machine learning techniques are also being considered as one of the methods for the identification and energy reconstruction of primary neutrons.

\begin{figure}[htbp]
  \centering
  \includegraphics[width=\textwidth]{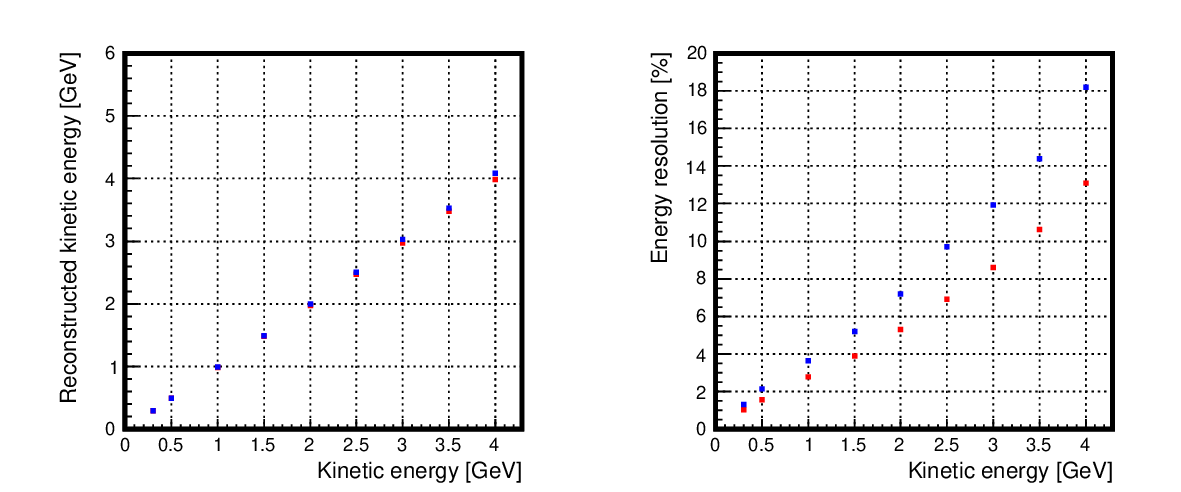}
  \caption{Left: The dependence of reconstructed kinetic energy on neutron energy for a cell time resolution of 100 ps (red) and 150 ps (blue). Right: The dependence of neutron energy resolution on neutron kinetic energy.}
  \label{fig:reconstruct}
\end{figure}

The dependence of neutron reconstructed energy and energy resolution is shown in Fig. \ref{fig:reconstruct} as a function of primary neutron energy. Results are shown for two cases of cell time resolutions, 100 ps (red) and 150 ps (blue). The simulated and reconstructed neutron energies are in agreement. For a time resolution of 100 ps, the neutron energy resolution is about 1\% for 300 MeV and increases up to 13\% for 4 GeV neutrons.

According to simulation results, about $10^9$ primary neutrons can be accumulated with the HGND during one month of operation at the BM@N. This estimation has been done for a beam rate of $5\times10^5$ per second, an efficiency of accelerator operation of 70\%, a target interaction length of 2\%, a mean primary neutron multiplicity of 0.1 neutrons per one interaction, and a mean efficiency of the HGND of 50\%. This number of neutrons will be sufficient to study the rapidity and transverse momentum dependencies of the direct and elliptic neutron flow with good precision.

\section{Summary and Outlook}
In this paper, the design and simulation results of the new compact HGND developed for neutron detection at the BM@N have been discussed. It was shown that the proposed time-of-flight HGND, assembled from layers of small cells with a time resolution of 100-150 ps and absorbers between these layers, will be able to identify neutrons and measure their energy in the range of 300 - 4000 MeV with an energy resolution of a few percent and an efficiency of about 50-60\%, depending on the neutron energy.

This new detector is planned to be constructed and used at the BM@N setup at the NICA complex, starting from 2026, to measure the azimuthal flow of neutrons in heavy-ion collisions at the BM@N energy range of 2-4 AGeV, where nuclear matter density of 2-4 $\rho_0$ is realized.

\section*{Acknowledgments}
This work was carried out at the Institute for Nuclear Research, Russian Academy of Sciences, and supported by the Russian Scientific Foundation grant \textnumero 22-12-00132. The authors thank S.Afanasiev (JINR) and A.Stavinskiy (Kurchatov Institute, NRC) for many helpful discussions and useful suggestions.

\bibliographystyle{elsarticle-num} 
\bibliography{references.bib}

\end{document}